\documentclass{article}
\usepackage{microtype}
\usepackage{graphicx}
\usepackage{subcaption}
\usepackage{booktabs} 

\usepackage{hyperref}

\usepackage[preprint]{icml2026}

\usepackage{amsmath}
\usepackage{mathtools}
\usepackage{amsthm}

\usepackage[capitalize,noabbrev]{cleveref}

\theoremstyle{plain}

\theoremstyle{definition}

\theoremstyle{remark}

\usepackage{xspace}
\usepackage{enumitem}
\usepackage{multirow}
\usepackage{placeins}
\usepackage{float}
\usepackage{xurl}

\begin{document}

\newcommand{\Comment}[1]{}

\newcommand{\rev}[1]{{\color{red}#1}}
\newcommand{\bhl}[1]{{\color{red}#1}}
\newcommand{\dep}[1]{{\color{red}#1}}

\newcommand{\todoc}[2]{{\textcolor{#1}{\textbf{#2}}}}
\definecolor{darkblue}{rgb}{0.0, 0.0, 0.55}

\newcommand{\todored}[1]{{\todoc{red}{\textbf{[[#1]]}}}}
\newcommand{\todogreen}[1]{\todoc{green}{\textbf{[[#1]]}}}
\newcommand{\todoblue}[1]{\todoc{blue}{\textbf{[[#1]]}}}
\newcommand{\todoorange}[1]{\todoc{orange}{\textbf{[[#1]]}}}
\newcommand{\todobrown}[1]{\todoc{brown}{\textbf{[[#1]]}}}
\newcommand{\todogray}[1]{\todoc{gray}{\textbf{[[#1]]}}}
\newcommand{\todopink}[1]{\todoc{purple}{\textbf{[[#1]]}}}
\newcommand{\todoteal}[1]{\todoc{teal}{\textbf{[[#1]]}}}
\newcommand{\tododarkblue}[1]{\todoc{darkblue}{\textbf{[[#1]]}}}

\newcommand{\todo}[1]{\todored{TODO: #1}}
\newcommand{\osditodo}[1]{\todoorange{TODO: #1}}

\newcommand{\para}[1]{\smallskip\noindent {\bf #1} }


\newcommand{\finding}[1]{
\begin{mdframed}[linecolor=gray,roundcorner=12pt,backgroundcolor=gray!15,
linewidth=3pt,innerleftmargin=2pt, leftmargin=0cm,rightmargin=0cm,
topline=false,bottomline=false,rightline = false]
#1
\end{mdframed}
}

\newcommand{\cellminipage}[2]{\begin{minipage}{#1}#2\end{minipage}}

\catcode`_=12 %
\renewcommand{\texttt}[1]{%
  \begingroup
  \ttfamily
  \begingroup\lccode`~=`/\lowercase{\endgroup\def~}{/\discretionary{}{}{}}%
  \begingroup\lccode`~=`[\lowercase{\endgroup\def~}{[\discretionary{}{}{}}%
  \begingroup\lccode`~=`.\lowercase{\endgroup\def~}{.\discretionary{}{}{}}%
  \begingroup\lccode`~=`_\lowercase{\endgroup\def~}{_\discretionary{}{}{}}%
  \catcode`/=\active\catcode`[=\active\catcode`.=\active\catcode`_=\active
  \scantokens{#1\noexpand}%
  \endgroup
}
\catcode`_=8 %

\newcommand{\lin}[1]{\todoblue{Lin: #1}} 
\newcommand{\yongle}[1]{\todoteal{Yongle: #1}}
\newcommand{\shangshu}[1]{\todoorange{Shangshu: #1}}
\newcommand{\sruthi}[1]{\tododarkblue{Sruthi: #1}}

\newcommand{\tool}{\textsc{Kevlar\-Flow}\xspace}

\newcommand{\BestAvgLatencyImprovement}{3.1x\xspace}
\newcommand{\BestPNNLatencyImprovement}{2.8x\xspace}
\newcommand{\BestAvgTTFTImprovement}{378.9x\xspace}
\newcommand{\BestPNNTTFTImprovement}{574.6x\xspace}
\newcommand{\MTTRImprovement}{20x\xspace}
\newcommand{\OurMTTRSec}{30\xspace}

\twocolumn[
  \icmltitle{Towards Resiliency in Large Language Model Serving with \tool}



  \icmlsetsymbol{equal}{*}

  \begin{icmlauthorlist}
    \icmlauthor{Shangshu Qian}{purdue}
    \icmlauthor{Kipling Liu}{purdue}
    \icmlauthor{P. C. Sruthi}{purdue}
    \icmlauthor{Lin Tan}{purdue}
    \icmlauthor{Yongle Zhang}{purdue}
  \end{icmlauthorlist}

  \icmlaffiliation{purdue}{Department of Computer Science, Purdue University, West Lafayette, IN, United States}

  \icmlcorrespondingauthor{Shangshu Qian}{shangshu@purdue.edu}
  \icmlcorrespondingauthor{P. C. Sruthi}{psruthi@purdue.edu}
  \icmlcorrespondingauthor{Lin Tan}{lintan@purdue.edu}
  \icmlcorrespondingauthor{Yongle Zhang}{yonglezh@purdue.edu}

  \icmlkeywords{LLM Serving, Fault Tolerance}

  \vskip 0.3in
]



\printAffiliationsAndNotice{}  


\begin{abstract}
Large Language Model (LLM) serving systems remain fundamentally fragile, where frequent hardware faults in hyperscale clusters trigger disproportionate service outages in the software stack. Current recovery mechanisms are prohibitively slow, often requiring up to 10 minutes to reinitialize resources and reload massive model weights. We introduce \tool, a fault tolerant serving architecture designed to bridge the gap between hardware unreliability and service availability. \tool leverages 1) decoupled model parallelism initialization, 2) dynamic traffic rerouting, and 3) background KV cache replication to maintain high throughput during partial failures. Our evaluation demonstrates that \tool reduces mean-time-to-recovery (MTTR) by \MTTRImprovement and, under failure conditions, improves average latency by \BestAvgLatencyImprovement, 99th percentile (p99) latency by \BestPNNLatencyImprovement, average time-to-first-token (TTFT) by \BestAvgTTFTImprovement, and p99 TTFT by \BestPNNTTFTImprovement with negligible runtime overhead in comparison to state-of-the-art LLM serving systems. 
\end{abstract}
\section{Introduction}
\label{sec:intro}

LLMs have become an essential component of modern digital infrastructure, demonstrating exceptional capabilities across a wide range of tasks. However, despite their ubiquity, the infrastructure supporting LLM serving remains fundamentally fragile~\cite{chu2025empirical}. Operational data reveals that outages are a common occurrence even for industry leaders, with significant service interruptions reported for platforms such as OpenAI~\cite{lu2024operational} and Anthropic’s Claude~\cite{bai2025understanding}. As these models are increasingly integrated into critical applications, the reliability of the underlying serving systems has emerged as a paramount concern.

\begin{figure}
    \centering
    \includegraphics[width=\linewidth]{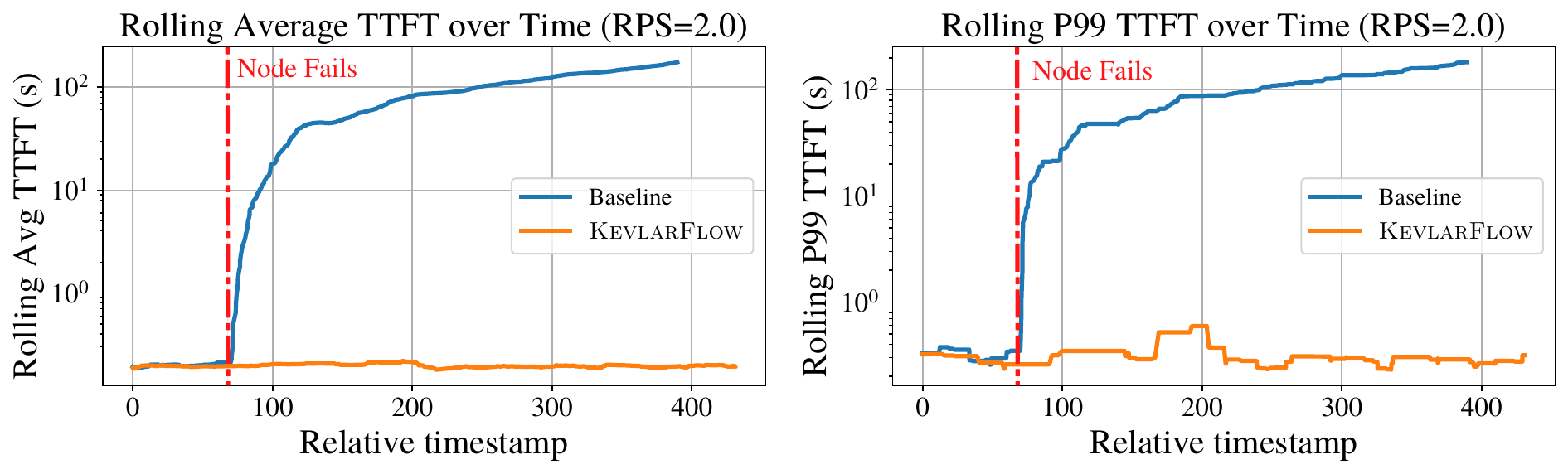}
    \caption{Rolling average and p99 TTFT comparison between a typical LLM serving framework (baseline) and \tool 
    under 2 requests per second (RPS). The red line indicates the time when one node fails. Y-axis is in log scale.}
    \label{fig:intro-ttft-under-failure}
\end{figure}

The root cause of this reliability crisis is threefold. 
First, \textbf{hardware failures} are common in production‑level GPU clusters~\cite{cui2025story, grattafiori2024llama} and frequently lead to program crashes~\cite{tiwari2015understanding}. 
Second, \textbf{the software stack} lacks robust fault tolerance mechanisms, further exacerbating the hardware instability.
Modern LLM serving frameworks (e.g., vLLM~\cite{kwon2023efficient} and TensorRT-LLM~\cite{trtllm}) rely on collective communication libraries such as Message Passing Interface (MPI)~\cite{forum1994mpi} and NVIDIA Collective Communication Library (NCCL) to implement complex model parallelism strategies~\cite{shoeybi2019megatron} (e.g., tensor and pipeline parallelism). Such communication libraries often optimize for throughput instead of fault tolerance~\cite{hu2025demystifying, weingram2023xccl} and create complex inter-dependencies among participating nodes since the initialization of the serving program. As a result, the failure of a single node will render all other model parallelism participants, while still alive, unserviceable for user requests. \Cref{fig:intro-ttft-under-failure} (blue lines) shows the rolling average and p99 TTFT of a typical LLM serving cluster under node failures. When one node fails, incoming requests pile up the queue quickly due to reduced serving capacity, increasing the TTFT and affecting user experience.

Third, \textbf{slow recovery} of LLM serving systems compounds these software and hardware issues by prolonging service degradation after a failure.
Restoring a failed instance typically involves 1) a complete re-initialization of serving node (e.g., one instance on Amazon EC2), and 2) the reloading of massive model weights from remote storage. This process can take as long as \emph{10 minutes}~\cite{jaiswal2025serving}, and the LLM service (partially) backed by the failed node will remain \emph{unresponsive} during the entire recovery time. Furthermore, widely used redundancy techniques from traditional distributed systems, such as data replication, have not been seamlessly adapted for the memory-intensive and state-heavy nature of LLM inference~\cite{strati2024dejavu}. While replication ensures high availability in databases, applying it to LLM serving remaining a challenging problem due to the unique constraints of GPU memory and the transient nature of the inference state.

To address these challenges, we introduce \tool, a fault tolerant serving architecture designed to bridge the gap between hardware unreliability and service availability. To solve the hardware instability, \tool draws insight from the inherent redundancy found within the load balancer groups of serving systems~\cite{jain2025performance}. By treating the cluster not as a collection of rigid, independent instances but as a flexible pool of resources, \tool enables high-throughput serving that degrades gracefully rather than failing catastrophically.

\tool achieves resiliency in LLM serving through three novel design contributions:

\begin{enumerate}[wide, labelwidth=!, labelindent=\parindent, topsep=0pt]

\item
\textbf{Dynamic Traffic Rerouting and Partial Availability.}
\tool introduces a traffic management strategy that handles partial model parallelism failures -- scenarios where only a subset of nodes in a model parallel group fails. Instead of marking the entire instance as offline, \tool dynamically reroutes user traffic around the failed node to other healthy instances in the load balancing group. Crucially, it keeps the remaining healthy parts of the damaged model in service, allowing failed nodes to be replaced in the background. This approach significantly improves system throughput under failure conditions by preventing the waste of functional GPU resources blocked by dependencies.

\item 
\textbf{Decoupled Model Parallelism Initialization.}
Unlike other frameworks, \tool decouples the initialization of model parallelism communicator from model weight loading, instead of performing both tasks at the start of the service. This architectural separation allows for dynamic reconfigurations of the parallelism schemes at runtime. If a node fails, the system can re-establish communication among the remaining healthy nodes without incurring the heavy overhead of a full restart or weight reload. 

\item
\textbf{Background KV Cache Replication.}
To mitigate the latency penalty of request restarting, \tool implements a low-overhead background replication mechanism for the KV cache. By replicating the intermediate inference state to other nodes in the load balancing group during runtime, \tool ensures that if a failure occurs, partially served requests can be continued near-instantly on a live node. This replication converts what would be a hard failure requiring a restart into a seamless migration, preserving the user's session context. 

\end{enumerate}

Our evaluation demonstrates that \tool effectively masks hardware failures with minimal performance penalties:

\begin{enumerate}[wide, labelwidth=!, labelindent=\parindent, topsep=0pt]
\item 
Under partial failure scenarios, \tool improves average latency by \BestAvgLatencyImprovement, p99 latency
by \BestPNNLatencyImprovement, average TTFT by \BestAvgTTFTImprovement, and p99 TTFT by \BestPNNTTFTImprovement compared to failure behaviors in state-of-the-art LLM serving frameworks.

\item 
\tool shortens the mean-time-to-recovery (MTTR) of node failures from 10 minutes~\cite{jaiswal2025serving} to \OurMTTRSec seconds, a \MTTRImprovement improvement.

\item 
The background KV cache replication incurs negligible overhead during normal operations, highlighting \tool's potential as an ``always-on'' fault tolerance solution for the next-generation AI infrastructure.

\end{enumerate}

\tool is the first LLM serving framework that can tolerate runtime node failures, allowing non-interruptive request handling during failure recovery.

\section{Related Work}

In this section, we introduce existing fault tolerance mechanisms in both LLM training and serving applications, and show that \tool is the only approach that can tolerate node failures in serving applications. 

\paragraph{Fault-Tolerant LLM Training.}

Communication and computation patterns are fundamentally different in LLM training and serving applications. 
Fault tolerance training techniques~\cite{gandhi:ReCycle, duan:Parcae} do not directly apply to serving for two reasons: 1) serving maintains transient states (i.e., KV cache) for each request between multiple forward passes, while training computes the model output in a single forward pass, and 2) serving has stricter latency requirements (e.g., TTFT) beyond throughput.

Existing checkpoint or migration based fault tolerance mechanisms~\cite{wang:GEMINI, jiang:MegaScale} in training cannot be directly applied in serving due to the lack of ability to capture KV cache. Moreover, even with checkpoints, ongoing requests are still paused during failure recovery, 
which exacerbates latency requirements potentially leading to SLO violations. 
Pipeline adaptation methods such as ReCycle~\cite{gandhi:ReCycle} rely on redirecting work into schedule bubbles. However, since LLM serving performs only a forward pass, it lacks the exploitable bubbles found in training schedules, rendering such techniques ineffective for inference fault tolerance.

\paragraph{Fault-Tolerant LLM Serving.}

Fault tolerance in LLM serving has evolved in two stages. The first is KV cache preservation through migration. DejaVu~\cite{strati2024dejavu} streams KV cache to CPU or remote memory for fast recovery, whereas SpotServe~\cite{miao:SpotServe} migrates state during preemption grace periods ($\sim$30 s). However, both approaches require restarting the serving instance after topology changes (i.e., recreating communicators and reloading state) which hurts latency and TTFT.

Recent efforts in this area focus on fault-tolerant communicators. 
AnchorTP~\cite{xu2025anchortp} preserves inference states via daemons for elastic tensor parallelism, but only tolerates intra-node GPU failures and requires weight migration between GPUs with KV cache recomputation, which adds latency. R$^2$CCL~\cite{wang2025reliable} provides NIC failover by utilizing multi-NIC hardware, but only tolerates NIC failures instead of node failures. Both mechanisms share a fundamental limitation. They treat the LLM serving instance as \emph{a single fault domain}, and cannot tolerate the failure of the serving node itself.

In contrast, \tool splits each LLM serving instance into multiple fault domains (i.e., independent nodes), and 
achieves continuous request handling after node failures through decoupled initialization, which enables re-establishing serving instances 
with existing healthy nodes in the load balancing group, maximizing the resource utilization, and minimizing the service downtime needed for either reloading/migrating model weights or re-provisioning additional resources in existing approaches.

\textbf{Other LLM serving systems} focus on efficiency rather than fault tolerance. Dynamic scheduling approaches~\cite{wu2024dlora, fu2024serverlessllm, sun:Llumnix} and prefill-decode disaggregation~\cite{zhong2024distserve, agrawal:Sarathi} optimize resource utilization and latency but do not focus on fault tolerance.

\section{Design of \tool}
\label{sec:design}

\begin{figure*}[]

\centering
\includegraphics[width=0.99\linewidth]{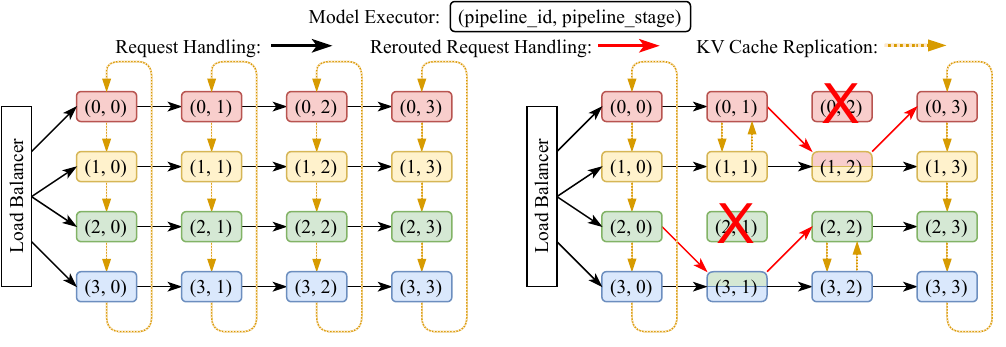}
\begin{subfigure}[t]{0.45\linewidth}
        \centering
        \vspace{-18pt}
        \caption{\tool under normal conditions.}
        \label{fig:design:normal}
    \end{subfigure}%
    \hfill
    \begin{subfigure}[t]{0.45\linewidth}
        \centering
        \vspace{-18pt}
        \caption{\tool under two node failures.}
        \label{fig:design:failure}
    \end{subfigure}

\caption{Design of \tool. The example shows a 4-stage pipeline parallelism serving (i.e., the model weights are split to four model executors). The load balancing group has four active model instances. A red ``X'' indicate a failed node.}
\label{fig:design}

\end{figure*}

Current LLM serving infrastructure suffers from a mismatch between the stochastic nature of hardware and the immutability assumptions in the software stack. While GPU clusters face high-frequency component failures~\cite{grattafiori2024llama}, software abstractions typically treat the cluster as immutable. \tool addresses this by abandoning the traditional ``fail-stop'' paradigm for a ``fail-stutter'' fault tolerance model~\cite{arpaci2001fail}. 
Using novel mechanisms of 1) decoupled initialization, 2) dynamic traffic rerouting, and 3) continuous state replication (\Cref{sec:design:ours}), \tool decouples logical availability from physical node health, transforming LLM serving infrastructures into a self-healing fabric.

\subsection{Background and Motivation}
\label{sec:design:background}

LLM serving systems generally employ a controller-worker architecture, comprising a centralized request metadata handler and multiple model executors. These executors are mapped to \emph{ranks} in a communication group (i.e., communicator) managed by libraries such as MPI or NCCL. In model parallelism deployments, the metadata handler often resides on the first rank. The initialization process of these serving instances generally follows a three-step sequence:

\begin{enumerate}[wide, labelwidth=!, labelindent=\parindent, topsep=0pt]

\item 
A \textbf{state sharing mechanism} is established to facilitate metadata exchange. This is commonly implemented via a PyTorch store (e.g., \texttt{TCPStore} in \texttt{torch.dis\-tributed})~\cite{li2020pytorch} or more sophisticated ones with MPI.

\item 
The system configures the \textbf{collective communicator} (e.g., an NCCL communicator) using the state sharing mechanism to define the topology for inter-device communications (e.g., \texttt{allreduce} operations).

\item 
The \textbf{model weights} are loaded, and the inference logic is configured to execute according to the model parallelism scheme~\cite{li2020pytorch}.

\end{enumerate}

While effective for performance, this architecture suffers from significant drawbacks regarding fault tolerance:

\begin{enumerate}[wide, labelwidth=!, labelindent=\parindent, topsep=0pt]

\item 
\textbf{Static Device Topology}: The device topology and communicator scope are fixed at startup~\cite{wang2025reliable}. For instance, \texttt{MPI_COMM_WORLD} is defined upon program launch and is immutable~\cite{Schulz_2023}. Adding or removing ranks from an existing communicator to handle failures or scaling is not fully supported~\cite{mixen56_2025}, often necessitating a full system restart instead.

\item
\textbf{Risks of Tensor Parallelism}: While intra-node tensor parallelism minimizes latency and maximizes throughput~\cite{narayanan2021efficient, shoeybi2019megatron}, it introduces a \emph{single point of failure}: the node itself. A failure in power or network connectivity for the node brings down the entire tensor parallel group. Additionally, the reliance on proprietary high-bandwidth interconnects (e.g., NVLink and NVSwitch) enforces vendor lock-in and restricts accessibility for academic and smaller entities~\cite{hooker2021hardware}.

\end{enumerate}

\subsection{The \tool Solution}
\label{sec:design:ours}

To address the drawbacks of existing systems, \tool adopts two core design principles: 1) reduce single points of failure, and 2) minimize data and control dependencies between model parallel ranks. 

\tool leverages the existing redundancy of model weights in load balancing groups, where multiple replicas of the model are used to handle concurrent user traffic~\cite{jain2025performance}. 
We adopt the multi-node pipeline parallelism as our base model parallelism scheme, which has gained significant traction in recent years~\cite{zhang2025td, jiang2025thunderserve, guo2025gllm, lin2025serving, luo2025multiplexed, ma2024hpipe} due to its distinct advantages over tensor parallelism: it 1) creates multiple independent fault domains for isolation, 2) achieves high throughput with low bandwidth requirements, and 3) eliminates the need for specialized interconnect hardware.

As illustrated in \Cref{fig:design}, \tool enhances this baseline architecture with three novel components to ensure fault tolerance properties:

\begin{enumerate}[wide, labelwidth=!, labelindent=\parindent, topsep=0pt]
\item \textbf{Decoupled Model Parallelism Initialization.}
We redesign the initialization process of LLM serving with a ground-up approach. Instead of relying on a single master process to coordinate a static global state, we use a decoupled model parallelism initialization process, where participating nodes autonomously connect to one another. The communication communicator is constructed only after nodes have been connected and verified as healthy.

This decoupling enables dynamic reconfiguration of the device topology. In the event of a node failure, the surviving nodes can rapidly identify a replacement node within the fault tolerance group and establish a new serving pipeline without tearing down the entire cluster.
\Cref{fig:design:failure} shows an example of this process. When node (0, 2) fails, node (0, 1) and (0, 3) quickly identifies another healthy node (1, 2), which holds the same portion of model weights as node (0, 2), to replace it. A new communicator and serving pipeline is formed promptly once the node failure is detected (more details below). 

\item \textbf{Dynamic Traffic Rerouting.}
Enabled by the decoupled initialization, \tool uses dynamic traffic rerouting when one pipeline is only partially available. In existing frameworks, a single node failure (e.g., node (0, 2) in \Cref{fig:design:failure}) interrupts the entire communicator, rendering all associated healthy GPUs (e.g., node (0, 0), (0, 1), and (0, 3)) idle and causing a total loss of that instance's serving capacity.

On the contrary, \tool contains the failure locally. When a node failure happens, \tool dynamically reroutes user traffic around the failed node (i.e., red arrows in \Cref{fig:design:failure}) through the newly created communicator. This ensures that the capacity drop is limited strictly to the failed node, while all other healthy nodes continue to process requests.

\item \textbf{Background KV Cache Replication.}
Current LLM serving systems utilize a KV cache to accelerate inference by storing attention states. Such data is usually stored in GPU memory or swapped to CPU memory under CUDA unified memory~\cite{NVidia_unified_memory}, and will be lost once the node failure happens, forcing a retry of the entire request.

To enable non-interruptive service, \tool replicates KV cache for each request to the GPU memory of other nodes in the load balancing group (e.g., yellow arrows in \Cref{fig:design:normal}). When failure occurs, a new serving pipeline involving the replication target (e.g., node (1, 2)) is established. In-progress requests will be served continuously on the replication target from the replicated state, avoiding retries and significantly reducing tail latency under failure conditions.

When node failure happens and the system is in a degraded state, KV cache replication targets will be automatically adjusted to exclude the nodes under traffic rerouting. For example, nodes (0, 2), (1, 2), (2, 1), and (3, 1) in \Cref{fig:design:failure} are excluded from KV cache replication. And the replication targets of nodes (1, 1) and (3, 2) are adjusted to other healthy nodes.

To avoid interference with request handling, \tool uses a block representation of KV cache~\cite{kwon2023efficient} and replicate it block-by-block in the background. A separate CUDA stream is used to overlap the communication with computation.

\end{enumerate}

A common concern is whether sufficient GPU memory exists to handle both the rerouted traffic and replication KV cache. Analyses of production workload traces~\cite{patke2025hierarchical, huang2024enova, jaiswal2025sageserve, zhang2025blitzscale} show that GPU is rarely saturated during serving, often fluctuating around 50\% - 60\% utilization to preserve memory headroom to ensure service level objectives (SLOs).

\tool utilizes such memory headroom to temporarily handle rerouted traffic and the replicated KV cache. When memory pressure happens, \tool drops the replicated KV cache and recomputes them if needed.

\subsection{Implementation Details}
\label{sec:design:impl}

\tool is implemented on top of TensorRT-LLM's PyTorch backend~\cite{trtllm}, a high performance LLM serving framework, providing an OpenAI-compatible server endpoint. Each node has a gRPC-based~\cite{grpc} RPC endpoint to facilitate inter-node communication (e.g., the formation of pipeline communicator). 

We port TensorRT-LLM to MPICH~\cite{gropp1996user} instead of using its original dependency of OpenMPI~\cite{gabriel2004open} due to MPICH's robust implementation of \texttt{MPI_Open_port}, \texttt{MPI_Comm_connect}, and \texttt{MPI_Intercomm_merge} functionalities. All are crucial for \tool's implementation of decoupled model parallelism initialization and traffic rerouting. 

\tool use NCCL to implement KV cache replication, enabling direct GPU-to-GPU replication when GPUDirect RDMA links are available. A distributed lock~\cite{hastings1990distributed} is implemented with PyTorch's TCPStore~\cite{li2020pytorch} to avoid deadlocks in the ring-shaped KV cache replication scheme (\Cref{fig:design:normal}) due to limitations in the NCCL' s \texttt{send} and \texttt{recv} functions.

\section{Evaluation}

We evaluate \tool using two virtual clusters rented from industry-leading cloud providers. One cluster is equipped with eight server nodes, and the other with 16 nodes. Each node has one NVIDIA A10 Tensor Core GPU with 24GB GDDR6 memory. Nodes within each cluster are distributed across four different datacenter locations in the eastern, central, western, and southern parts of the United States, simulating a geo-distributed load balancing group.

Each node is connected to Internet through a 1Gbps Ethernet port, and does not have any specialized interconnection hardware (e.g., InfiniBand). Nodes in different datacenters are in different autonomous systems (AS), connected through commercial Internet transit providers (e.g., Arelion, Level 3, and GTT), simulating \emph{affordable} computing resources available to academic research groups.

We simulate requests to LLM serving systems using requests from ShareGPT~\cite{ShareGPT}, a popular~\cite{gao2025apt, fu2024serverlessllm, wang2025burstgpt, xiang2025aegaeon, song2026xy, gong2025past, zhao2024alise, he2025resource, wu2024dlora, patke2024queue, papaioannou2024importance, zhong2024distserve, feng2025windserve} dataset evaluating the performance of LLM serving systems. We simulate the arrival time of requests using Poisson distribution under different parameters of request rate (RPS, or request per second).

We use Llama-3.1-8B model~\cite{llama31} with a 4-stage pipeline parallelism on our serving instances. Each pipeline stage is served on one node in the virtual cluster. We place each model instance on four nodes located in the same datacenter. In the first virtual cluster with eight nodes, the load balancing group has two model instances. In the second virtual cluster with 16 nodes, the load balancing group has four model instances. The load balancer distributes requests evenly across all instances in the load balancing group.

To evaluate \tool's performance, we focus on answering the following questions:
\begin{enumerate}[wide, labelwidth=!, labelindent=\parindent, topsep=0pt, noitemsep]
\item How does \tool perform under node failures?
\item How much improvement does \tool have on failure recovery time?
\item How much overhead does \tool introduce?
\end{enumerate}

\subsection{Baseline Performance}

\begin{figure}[tb]
    \centering

    \includegraphics[width=\linewidth]{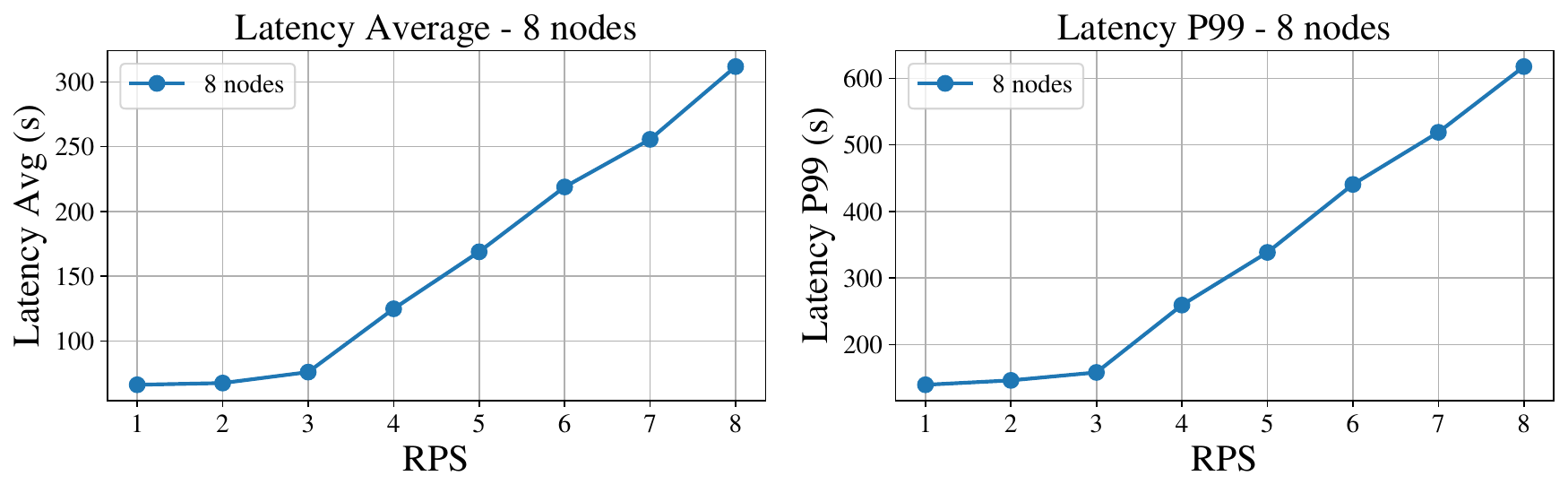}
    \includegraphics[width=\linewidth]{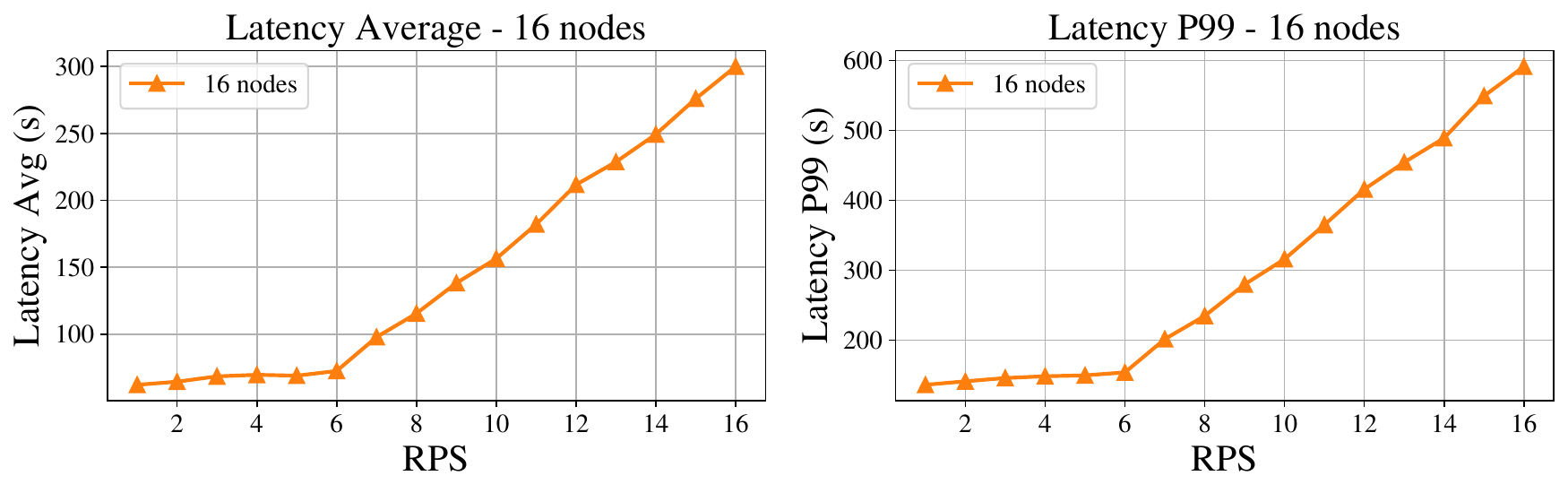}

    \caption{Baseline latency performance of TensorRT-LLM. The performance of 8-node cluster is marked with blue dots, and that of the 16-node cluster is marked with orange triangles. Figures on the left are for the average latency, and figures on the right are for the p99 latency.}
    \label{fig:trtllm-baseline-latency}
\end{figure}

\begin{figure}[tb]
    \centering

    \includegraphics[width=\linewidth]{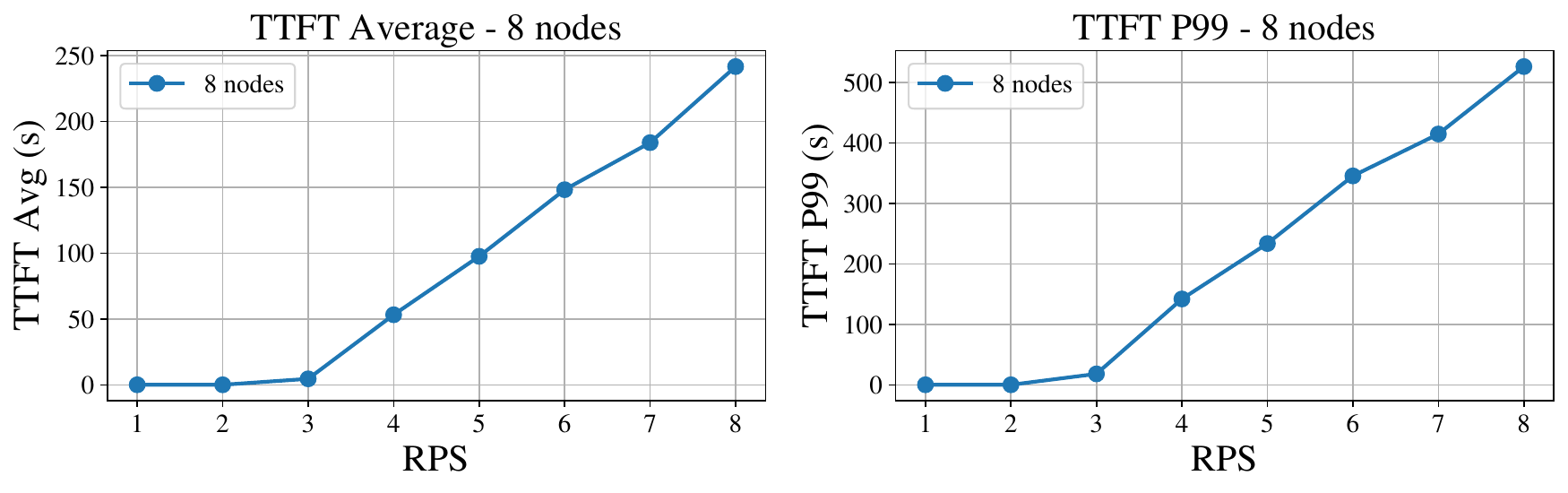}
    \includegraphics[width=\linewidth]{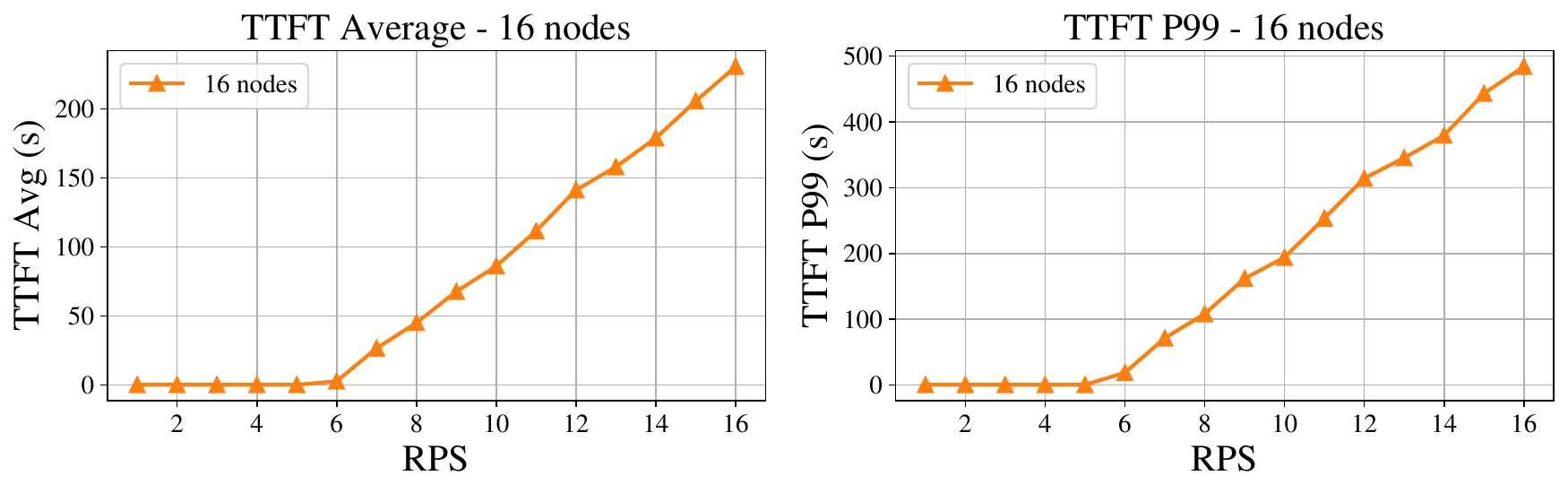}

    \caption{Baseline TTFT of TensorRT-LLM. The performance of 8-node cluster is marked with blue dots, and that of the 16-node cluster is marked with orange triangles. Figures on the left are for the average TTFT, and figures on the right are for the p99 TTFT.}
    \label{fig:trtllm-baseline-ttft}
\end{figure}

We first establish the baseline performance of TensorRT-LLM on our virtual clusters. \Cref{fig:trtllm-baseline-latency} shows the baseline latency of TensorRT-LLM serving Llama-3.1-8B with a 4-stage pipeline parallelism. The figure shows both the average and p99 latency on the 8-node and 16-node cluster. In the 8-node cluster, request latency starts to grow when the RPS (request/sec) increases from 3 to 4, indicating a significant growth request queue. Similar transition exists in the 16-node cluster when RPS grows from 6 to 7. This is expected as the 16-node cluster doubles the capacity of the 8-node cluster.

Time-to-first-token (TTFT), an important measurement for the service level objectives (SLOs)~\cite{hong2025sola}, shows a similar trend as the latency. As shown in \Cref{fig:trtllm-baseline-ttft}, requests start to queue up at a RPS of 3 in the 8-node cluster and at a RPS of 6 in the 16-node cluster.

Time-per-output-token (TPOT), a metric measuring the per-token latency of LLM serving instances, remains nearly constant in all baseline benchmark scenarios, with an average TPOT of 163ms/token and a p99 TPOT of 203ms/token. This is the behavior of TensorRT-LLM's default batch scheduler, and we use it in all our benchmark to ensure a fair comparison.

\subsection{Performance Under Node Failure}
\label{sec:eval:perf-under-failure}

\begin{figure*}[tb]
\centering

\begin{subfigure}{0.49\linewidth}
\includegraphics[width=\linewidth]{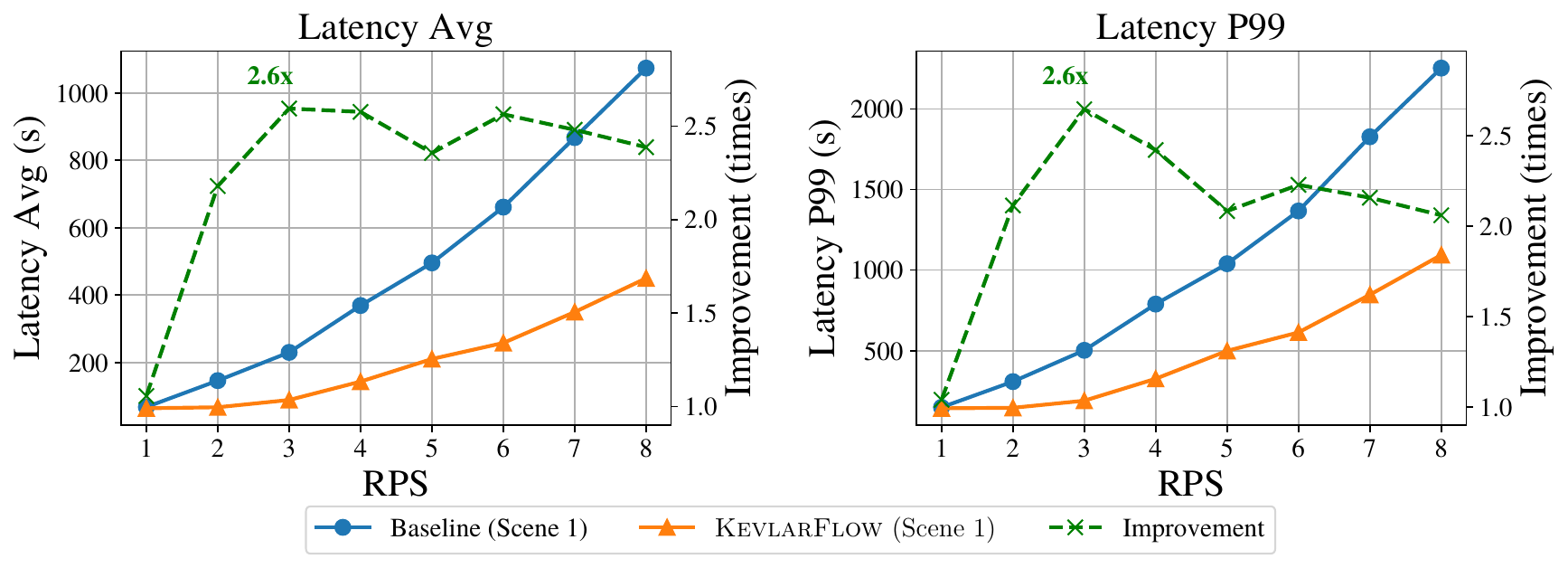}
\caption{Failure scenario 1 latency.}
\end{subfigure}
\hfill
\begin{subfigure}{0.49\linewidth}
\includegraphics[width=\linewidth]{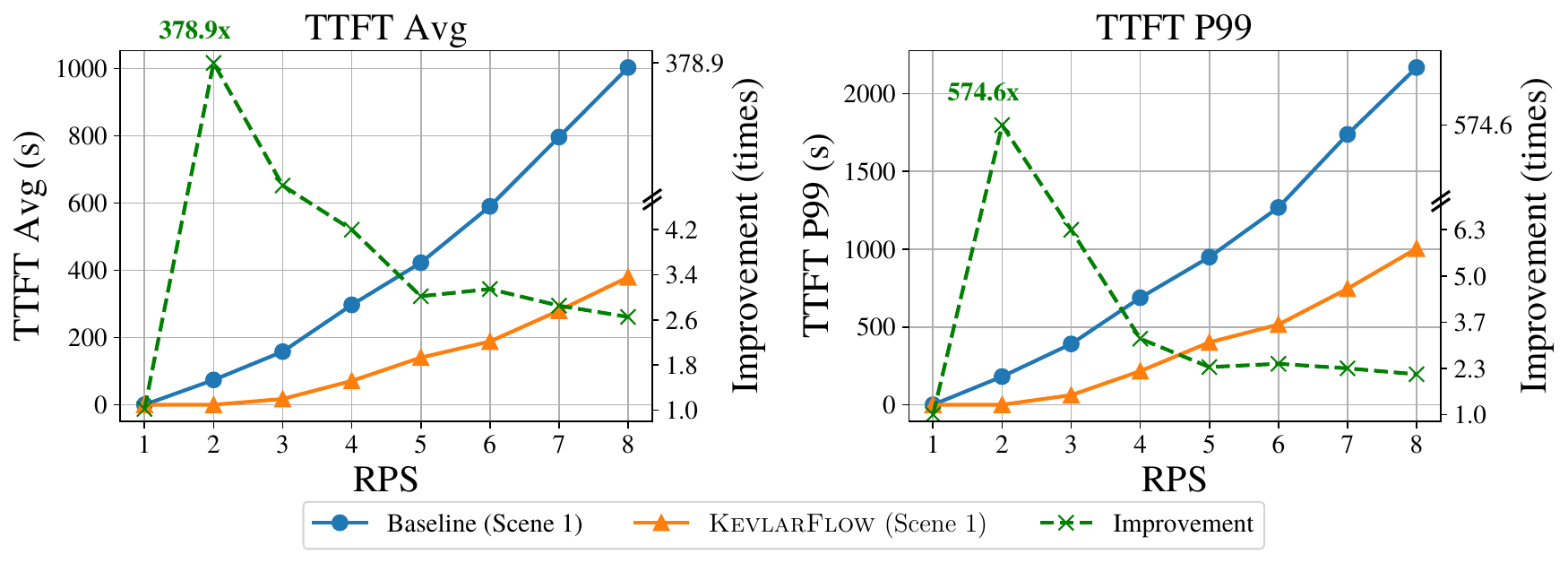}
\caption{Failure scenario 1 TTFT.}
\end{subfigure}

\begin{subfigure}{0.49\linewidth}
\includegraphics[width=\linewidth]{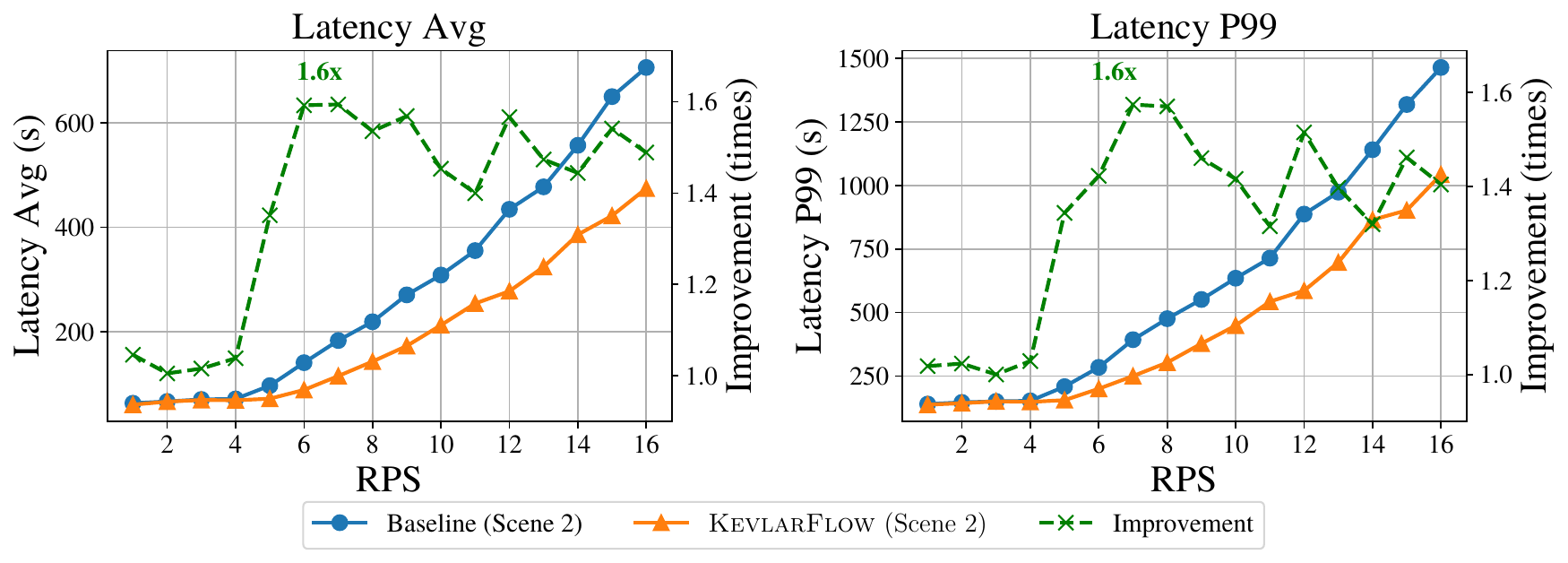}
\caption{Failure scenario 2 latency.}
\end{subfigure}
\hfill
\begin{subfigure}{0.49\linewidth}
\includegraphics[width=\linewidth]{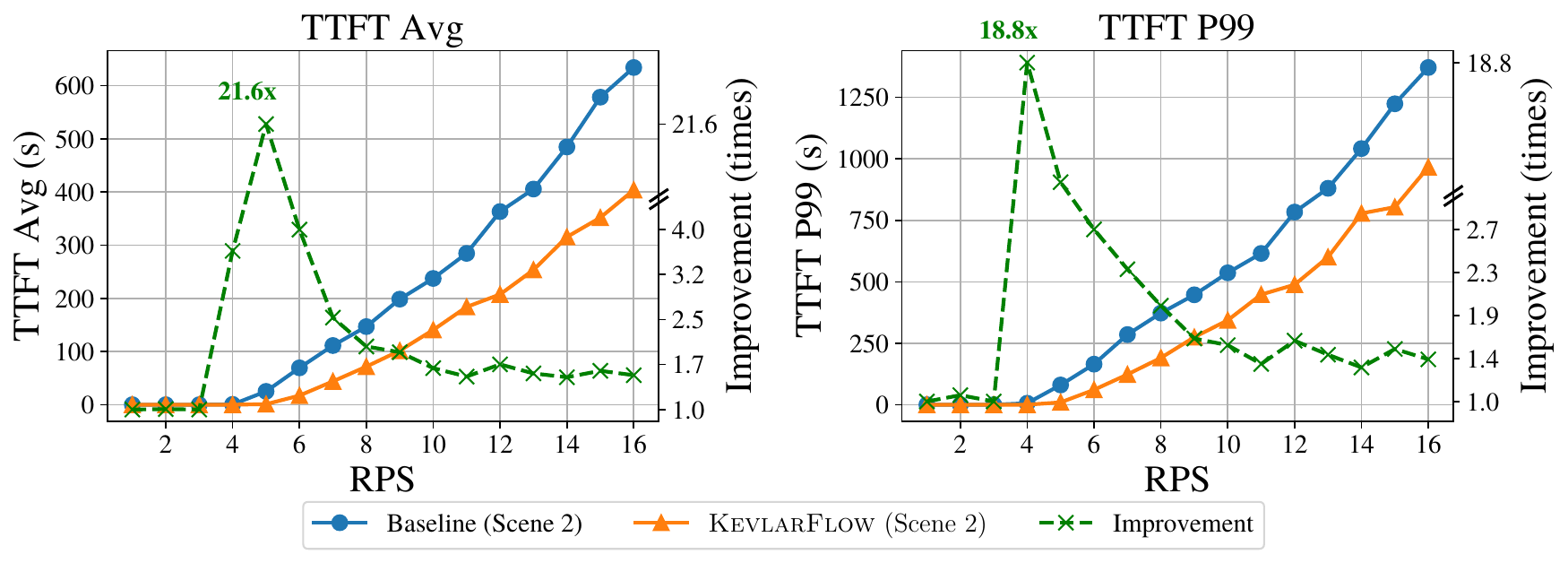}
\caption{Failure scenario 2 TTFT.}
\end{subfigure}

\begin{subfigure}{0.49\linewidth}
\includegraphics[width=\linewidth]{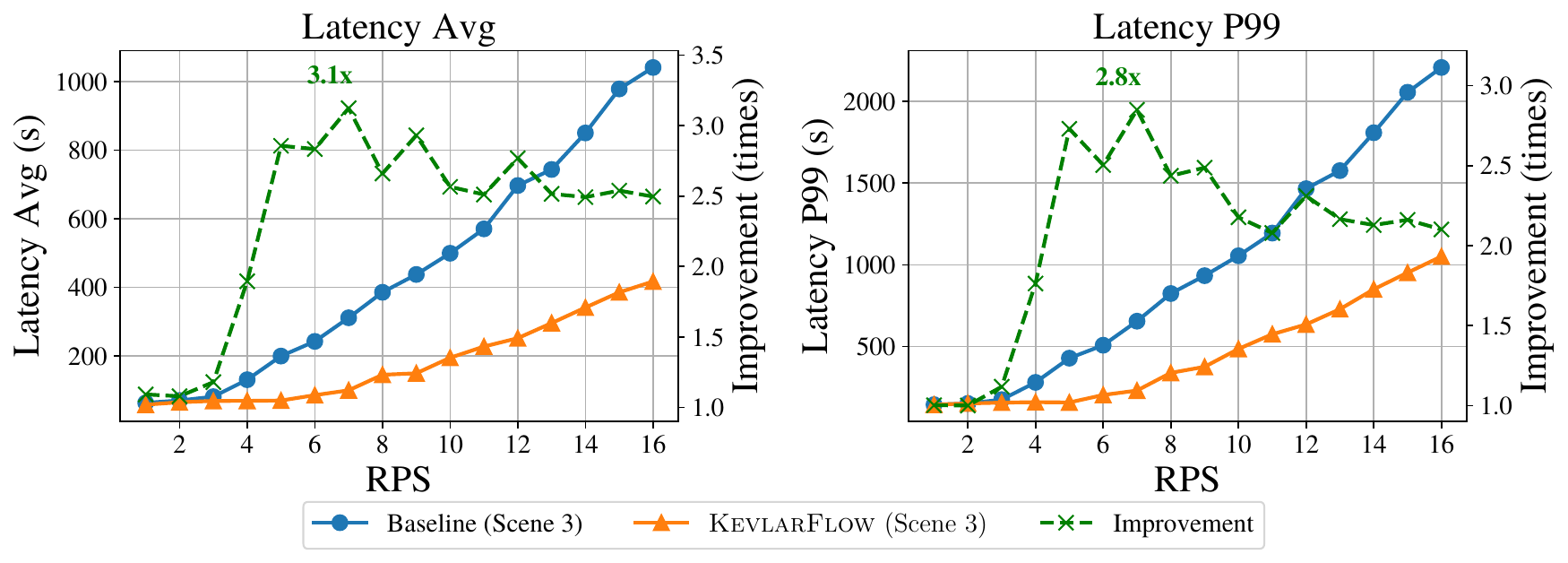}
\caption{Failure scenario 3 latency.}
\end{subfigure}
\hfill 
\begin{subfigure}{0.49\linewidth}
\includegraphics[width=\linewidth]{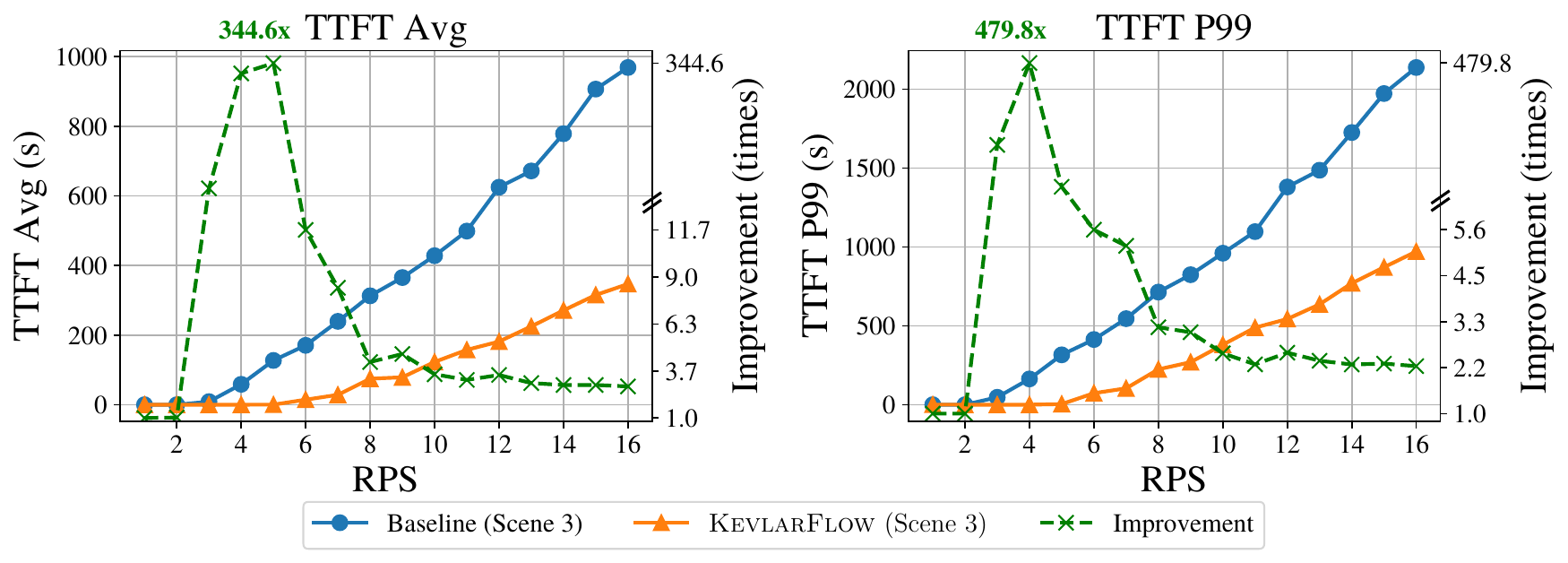}
\caption{Failure scenario 3 TTFT.}
\end{subfigure}

\caption{Comparison of \tool and baseline under three node failure scenarios. Performance of the standard fault behavior is plotted using blue dots, \tool's performance is plotted in orange triangles. The improvement of \tool over the standard fault behavior is plotted in green dotted lines. The highest improvement is annotated in each figure. }
\label{fig:perf-under-failure}
\end{figure*}

To demonstrate \tool's fault tolerance capabilities, we evaluate \tool under node failures against the standard fault behavior (\Cref{sec:intro}, \Cref{sec:design:ours}): when one node in the serving pipeline fails, the entire pipeline ceases to serve user requests. Under standard fault behavior, the load balancer will distribute incoming requests evenly to the remaining healthy pipelines. Any in-progress requests will be immediately retried once the node failure is detected. 

\textbf{Generality of the standard fault behavior.}
As discussed in \Cref{sec:intro} and \Cref{sec:design:background}, existing collective communication libraries such as MPI and NCCL assume a static device topology, and state-of-the-art serving frameworks (e.g., TensorRT-LLM, vLLM) strictly adhere to this assumption. \tool distinguishes itself as the first LLM serving framework capable of tolerating runtime node failures without service interruption. In contrast, all existing fault-tolerant approaches, including DejaVu~\cite{strati2024dejavu}, AnchorTP~\cite{xu2025anchortp}, and R$^2$CCL~\cite{wang2025reliable}, would follow the standard fault behavior. Thus, the comparative evaluation presented in this section is applicable to all such systems.

We compare \tool against standard fault behavior under three scenarios:

\begin{enumerate}[wide, labelwidth=!, labelindent=\parindent, topsep=0pt]
\item In the 8-node cluster, we inject failure to one node, failing one pipeline in the load balancing group.
\item In the 16-node cluster, we inject failure to one node, failing one pipeline in the load balancing group.
\item In the 16-node cluster, we inject failures to two nodes, failing two pipelines in the load balancing group.
\end{enumerate}

\Cref{fig:perf-under-failure} shows the comparison of average latency, average TTFT, p99 latency, and p99 TTFT of \tool against the standard fault behavior under three different fault scenarios. \tool achieves an improvement in average latency, p99 latency, average TTFT, and p99 TTFT of up to \textbf{3.1x}, \textbf{2.8x}, \textbf{378.9x}, and \textbf{574.6x} respectively. Detailed data of the benchmark can be found in \Cref{tab:perf-under-failure} in the appendix. 

\begin{figure}
    \centering
    \includegraphics[width=\linewidth]{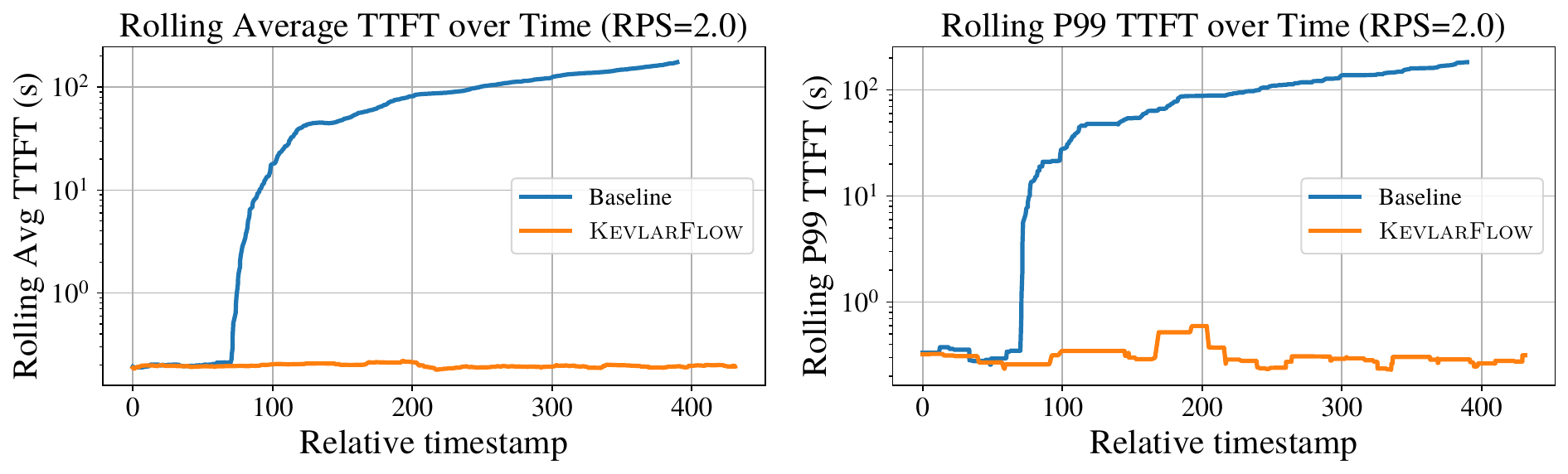}
    
    \caption{Rolling average and p99 TTFT over time under node failure scenario 1 (RPS=2.0). \tool is represented by the orange line, and standard fault behavior is represented by the blue line. Y-axis is in log scale.}
    \label{fig:rolling-ttft}
\end{figure}

\textbf{Explanation of the massive TTFT improvement.} 
In the event of a node failure, \tool utilizes dynamic traffic rerouting to keep the remaining healthy nodes of the degraded pipeline operational. This approach preserves a significantly larger fraction of the load balancing group's capacity compared to standard fault behavior, which often render an entire pipeline unusable upon a single failure. Consequently, \tool delays system saturation, allowing the scheduler to handle higher request rates without accumulating a backlog of queued requests.

\begin{figure}
    \centering
    \includegraphics[width=\linewidth]{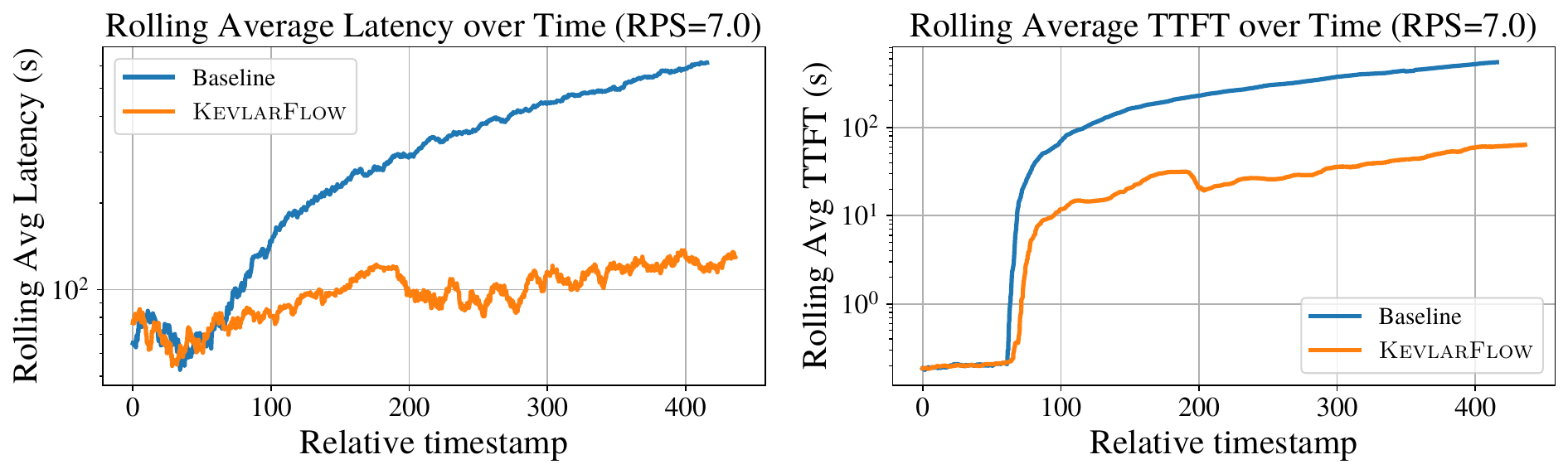}
    
    \caption{Rolling average latency and TTFT over time under node failure scenario 3 (RPS=7.0). \tool is represented by the orange line, and standard fault behavior is represented by the blue line. Y-axis is in log scale.}
    \label{fig:rolling-qps7}
\end{figure}

\Cref{fig:rolling-ttft} illustrates the rolling average and p99 TTFT for failure scenario 1 at an RPS of 2.0, where \tool demonstrates its most substantial improvement over the baseline. Upon node failure, serving clusters operating under standard fault behavior rapidly accumulate a backlog of queued requests, causing immediate spikes in the TTFT of incoming traffic. In contrast, \tool leverages dynamic traffic rerouting and KV cache replication to sustain uninterrupted request handling, thereby maintaining stable average and p99 TTFT. As shown in \Cref{fig:rolling-qps7}, this resilience persists even when the serving system is saturated (i.e., RPS $>3$ in scenario 1, and RPS $\ge 6$ in scenarios 2 and 3). \tool delivers significant performance gains over standard fault behavior -- achieving improvements of up to 3.18x, 2.71x, and 5.64x in scenarios 1, 2, and 3, respectively -- underscoring the versatility of \tool under diverse load conditions.

\subsection{Failure Recovery Time}

Current LLM serving systems suffer from prolonged error recovery times due to the heavy overhead of instance initialization and model weight loading. This process can take up to 10 minutes~\cite{jaiswal2025serving}.
During this downtime, the serving instance remains completely unresponsive to user requests.
Standard KV cache replication techniques, such as DejaVu~\cite{strati2024dejavu}, are ineffective at accelerating this specific recovery phase because they rely on the existence of a live serving instance to receive the state, which is unavailable during a hard node failure.

\tool addresses this by leveraging decoupled initialization and dynamic traffic rerouting to enable rapid service restoration. Upon detecting a node failure, \tool automatically: 1) locates a healthy alternative node within the load balancing group that holds the same model weights,
2) re-establishes the inter-node communicator,
and 3) resumes the processing of user requests. 
This recovery mechanism overlaps the lengthy full-instance initialization with a degraded yet functional serving pipeline, significantly reducing both latency and TTFT for user requests during failure events.

\begin{figure}
    \centering
    \includegraphics[trim={2mm 0 2mm 0}, width=\linewidth, clip]{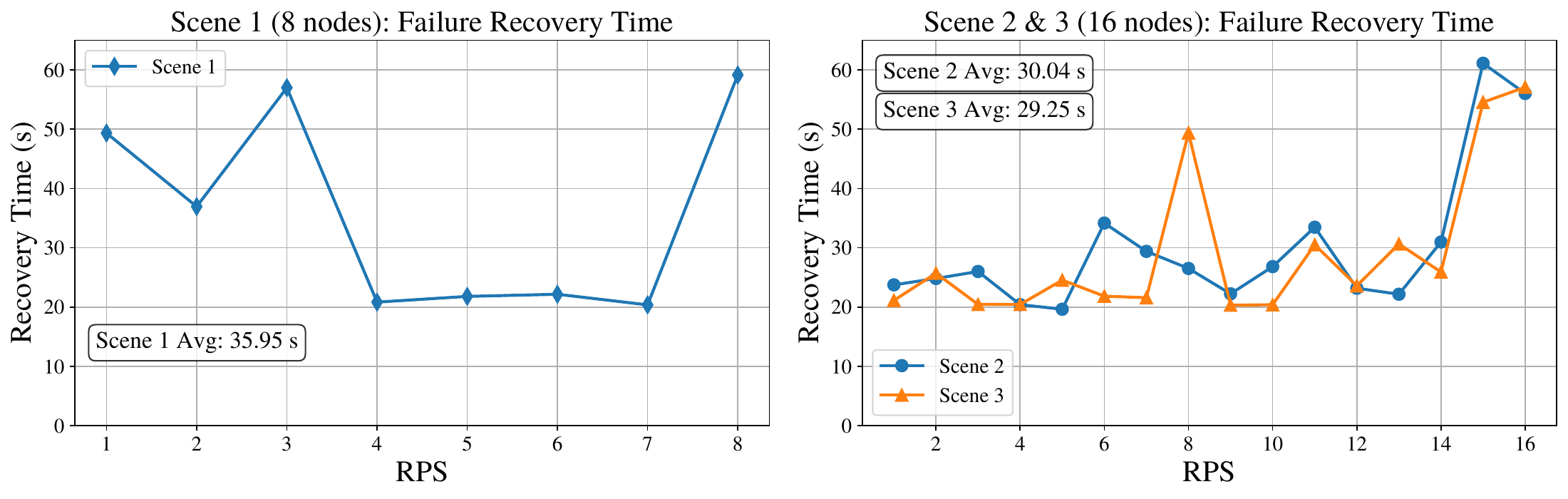}
    
    \caption{Failure recovery time of \tool under three failure scenarios.}
    \label{fig:failure-recovery}
\end{figure}

\Cref{fig:failure-recovery} illustrates the failure recovery time of \tool across the three scenarios defined in \Cref{sec:eval:perf-under-failure}. \tool achieves an average recovery time of 35s in scenario 1, 30s in scenario 2, and 29s in scenario 3. These results represent a \textbf{20x} improvement over the mean-time-to-recovery (MTTR) of state-of-the-art LLM serving systems.

Note that user requests continue to be processed throughout \tool's failure recovery phase. As shown in \Cref{fig:failure-recovery}, the recovery duration does not increase with the RPS. Instead, it fluctuates marginally around the average value. This stability demonstrates \tool's scalability and its ability to maintain consistent fault tolerance under varying workload intensities.

\textbf{Comparison with hot spare approaches.}
Another strategy for reducing failure recovery time is leveraging hot spare nodes~\cite{miao:SpotServe}, which remain online but idle during normal operations. While this allows for rapid replacement upon node failure, it inherently wastes valuable GPU resources. In contrast, \tool employs dynamic traffic rerouting to maintain continuous service using all available healthy nodes, while replacement resources are initialized in the background. Consequently, \tool offers superior cost-efficiency compared to hot spare approaches, as it eliminates the need to provision redundant, idle GPU capacity at system startup.

\subsection{Runtime Overhead}

\begin{figure}
    \centering
    \begin{subfigure}{\linewidth}
        \includegraphics[width=\linewidth]{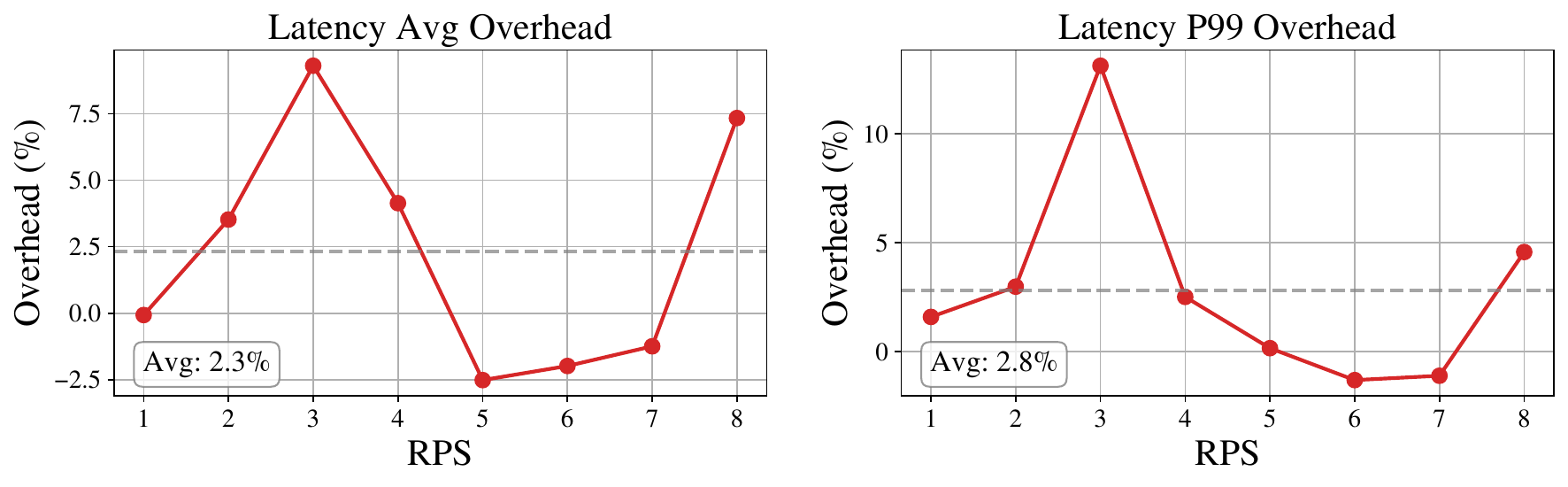}
        \caption{Overhead in the 8-node cluster.}
        \label{fig:overhead:8node}
    \end{subfigure}

    \begin{subfigure}{\linewidth}
        \includegraphics[width=\linewidth]{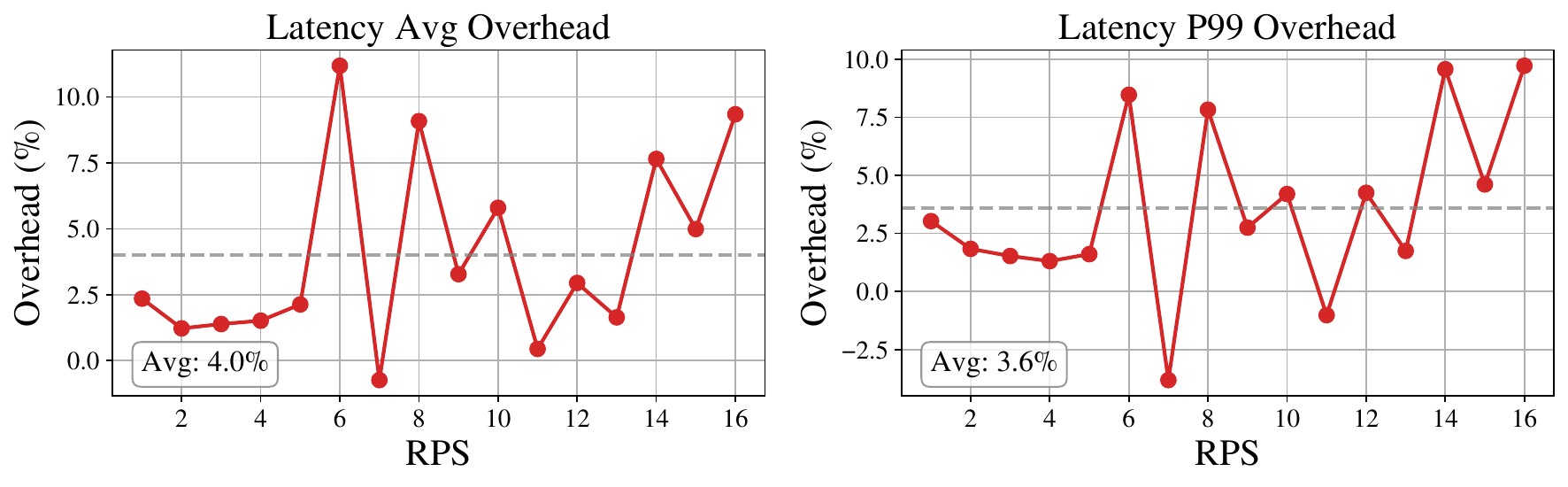}
        \caption{Overhead in the 16-node cluster.}
        \label{fig:overhead:16node}
    \end{subfigure}
    
    \caption{Runtime overhead of \tool. Some overhead values are negative due to non-determinism in the execution. We consider them as 0\% overhead in practice.}
    \label{fig:overhead}
\end{figure}

A critical requirement for fault-tolerant serving systems is minimizing the performance penalty imposed during normal, failure-free operation. We evaluate the overhead of \tool by measuring the impact of its continuous background KV cache replication on end-to-end serving latency.

As detailed in \Cref{sec:design:ours}, \tool mitigates resource contention by performing block-wise KV cache replication on a dedicated CUDA stream, effectively overlapping communication costs with computation. 

Our evaluation confirms that \tool incurs \textbf{minimal runtime overhead} compared to the baseline. As shown in \Cref{fig:overhead:8node}, the 8-node cluster exhibits an average latency overhead of only 2.3\% and a p99 overhead of 2.8\%. Similarly, \Cref{fig:overhead:16node} shows that on the larger 16-node cluster, the average and p99 overheads remain constrained to 4.0\% and 3.6\%, respectively. The overhead fluctuates across different RPS rates while occasionally dropping below zero due to non-determinism in execution. It confirms that the background replication does not introduce systematic contention. This low overhead highlights \tool's viability as an ``always-on'' fault tolerance solution for the hyper-scale AI infrastructure.
\section{Conclusion}

We present \tool, a fault tolerant LLM serving architecture that bridges the gap between unreliable hardware and reliable LLM services.
Our extensive evaluation on geo-distributed clusters demonstrates that \tool significantly enhances LLM serving systems' fault tolerant properties. It reduces the MTTR by \textbf{\MTTRImprovement} -- from 10 minutes to roughly \OurMTTRSec seconds. Furthermore, under partial failure conditions, \tool improves average and p99 TTFT by up to \textbf{\BestAvgTTFTImprovement} and \textbf{\BestPNNTTFTImprovement} respectively in comparison to state-of-the-art LLM serving systems with negligible runtime overhead during normal operations.

\section*{Acknowledgments}
The authors thank David Helsel for helping with the implementations.
This research is partially supported by NSF 1901242, 2006688, 2300562, and a Meta Research Award.
Any opinions, findings, and conclusions in this paper are those of the authors 
only and do not necessarily reflect the views of the sponsors.

\section*{Impact Statement}

This paper presents work whose goal is to advance the field of Machine
Learning. There are many potential societal consequences of our work, none
which we feel must be specifically highlighted here.

\bibliography{paper}
\bibliographystyle{icml2026}

\newpage
\appendix
\onecolumn
\section{Additional Results}

\begin{table*}[htb]
\centering
\caption{Comparison of \tool and baseline under three node failure scenarios. ``Base.'' indicates the standard fault behavior. ``Imp.'' means the improvement of \tool over the standard fault behavior. Improvement less than 1x is caused by non-determinism in execution.}

\resizebox{\textwidth}{!}{
\begin{tabular}{l|c|rrr|rrr|rrr|rrr}
\toprule 
 & & \multicolumn{3}{c|}{Latency Avg (s)} & \multicolumn{3}{c|}{TTFT Avg (s)} & \multicolumn{3}{c|}{Latency P99 (s)} & \multicolumn{3}{c}{TTFT P99 (s)} \\
Scene & RPS & Base. & Ours & Imp. & Base. & Ours & Imp. & Base. & Ours & Imp. & Base. & Ours & Imp. \\
\midrule
\multirow{8}{*}{\rotatebox[origin=c]{90}{Scene 1}} & 1.0 & 67.70 & 64.06 & 1.06x & 0.20 & 0.20 & 1.02x & 148.87 & 143.05 & 1.04x & 0.33 & 0.34 & 0.97x \\
 & 2.0 & 146.15 & 67.07 & 2.18x & 73.84 & 0.19 & 378.91x & 308.48 & 145.92 & 2.11x & 181.18 & 0.32 & 574.56x \\
 & 3.0 & 230.02 & 88.69 & 2.59x & 158.11 & 17.30 & 9.14x & 502.54 & 189.97 & 2.65x & 391.10 & 61.98 & 6.31x \\
 & 4.0 & 369.08 & 143.29 & 2.58x & 297.26 & 71.08 & 4.18x & 789.27 & 326.14 & 2.42x & 687.86 & 216.26 & 3.18x \\
 & 5.0 & 495.64 & 210.39 & 2.36x & 422.62 & 140.47 & 3.01x & 1039.56 & 499.11 & 2.08x & 948.84 & 400.99 & 2.37x \\
 & 6.0 & 661.22 & 258.00 & 2.56x & 590.29 & 188.27 & 3.14x & 1366.77 & 613.55 & 2.23x & 1267.99 & 515.41 & 2.46x \\
 & 7.0 & 867.55 & 349.84 & 2.48x & 796.53 & 280.19 & 2.84x & 1826.76 & 847.07 & 2.16x & 1737.31 & 745.76 & 2.33x \\
 & 8.0 & 1074.00 & 449.96 & 2.39x & 1003.40 & 379.47 & 2.64x & 2252.86 & 1093.81 & 2.06x & 2167.29 & 1003.85 & 2.16x \\
\midrule 
\multirow{16}{*}{\rotatebox[origin=c]{90}{Scene 2}} & 1.0 & 62.56 & 59.80 & 1.05x & 0.19 & 0.20 & 0.95x & 139.63 & 137.19 & 1.02x & 0.33 & 0.34 & 0.96x \\
 & 2.0 & 65.99 & 65.64 & 1.01x & 0.20 & 0.20 & 1.01x & 146.18 & 142.87 & 1.02x & 0.34 & 0.32 & 1.06x \\
 & 3.0 & 69.84 & 68.75 & 1.02x & 0.20 & 0.20 & 0.99x & 150.13 & 150.40 & 1.00x & 0.34 & 0.35 & 0.98x \\
 & 4.0 & 70.85 & 68.22 & 1.04x & 0.74 & 0.20 & 3.64x & 152.32 & 148.09 & 1.03x & 6.35 & 0.34 & 18.85x \\
 & 5.0 & 96.31 & 71.27 & 1.35x & 25.24 & 1.17 & 21.58x & 208.28 & 155.03 & 1.34x & 80.55 & 9.84 & 8.18x \\
 & 6.0 & 140.33 & 88.15 & 1.59x & 69.49 & 17.42 & 3.99x & 284.19 & 199.84 & 1.42x & 165.26 & 60.92 & 2.71x \\
 & 7.0 & 182.84 & 114.71 & 1.59x & 111.47 & 44.10 & 2.53x & 393.03 & 249.74 & 1.57x & 285.59 & 123.26 & 2.32x \\
 & 8.0 & 218.79 & 142.49 & 1.54x & 147.06 & 71.83 & 2.05x & 475.71 & 303.05 & 1.57x & 372.80 & 190.82 & 1.95x \\
 & 9.0 & 270.33 & 172.37 & 1.57x & 198.74 & 101.80 & 1.95x & 551.31 & 377.61 & 1.46x & 447.16 & 274.88 & 1.63x \\
 & 10.0 & 308.39 & 212.22 & 1.45x & 237.64 & 140.78 & 1.69x & 635.05 & 448.57 & 1.42x & 536.55 & 343.79 & 1.56x \\
 & 11.0 & 355.24 & 253.80 & 1.40x & 284.79 & 184.06 & 1.55x & 714.06 & 542.99 & 1.32x & 615.45 & 448.20 & 1.37x \\
 & 12.0 & 434.45 & 277.38 & 1.57x & 363.37 & 207.60 & 1.75x & 887.34 & 585.88 & 1.51x & 783.54 & 488.21 & 1.60x \\
 & 13.0 & 477.47 & 324.02 & 1.47x & 405.76 & 253.63 & 1.60x & 973.63 & 697.05 & 1.40x & 880.20 & 600.52 & 1.47x \\
 & 14.0 & 557.11 & 385.81 & 1.44x & 484.94 & 315.64 & 1.54x & 1140.89 & 864.77 & 1.32x & 1041.62 & 778.17 & 1.34x \\
 & 15.0 & 650.51 & 422.08 & 1.54x & 578.53 & 351.99 & 1.64x & 1318.74 & 902.43 & 1.46x & 1223.75 & 803.75 & 1.52x \\
 & 16.0 & 706.36 & 474.50 & 1.49x & 634.31 & 404.00 & 1.57x & 1464.48 & 1043.13 & 1.40x & 1371.26 & 966.13 & 1.42x \\
\midrule
\multirow{16}{*}{\rotatebox[origin=c]{90}{Scene 3}} & 1.0 & 63.05 & 57.71 & 1.09x & 0.19 & 0.20 & 0.95x & 143.49 & 143.22 & 1.00x & 0.32 & 0.70 & 0.45x \\
 & 2.0 & 69.96 & 64.81 & 1.08x & 0.20 & 0.19 & 1.03x & 150.65 & 150.89 & 1.00x & 0.31 & 0.32 & 0.98x \\
 & 3.0 & 81.06 & 68.72 & 1.18x & 8.96 & 0.20 & 44.96x & 174.10 & 155.77 & 1.12x & 47.22 & 0.34 & 139.02x \\
 & 4.0 & 130.50 & 68.85 & 1.90x & 58.77 & 0.20 & 291.69x & 279.86 & 158.76 & 1.76x & 163.65 & 0.34 & 479.81x \\
 & 5.0 & 199.43 & 69.81 & 2.86x & 127.32 & 0.37 & 344.57x & 428.10 & 156.81 & 2.73x & 316.37 & 4.29 & 73.78x \\
 & 6.0 & 242.23 & 85.48 & 2.83x & 170.63 & 14.62 & 11.67x & 507.31 & 202.63 & 2.50x & 413.60 & 73.28 & 5.64x \\
 & 7.0 & 311.04 & 99.58 & 3.12x & 239.55 & 28.56 & 8.39x & 654.54 & 229.75 & 2.85x & 545.80 & 104.22 & 5.24x \\
 & 8.0 & 385.38 & 144.88 & 2.66x & 312.91 & 74.90 & 4.18x & 823.90 & 338.17 & 2.44x & 714.20 & 224.51 & 3.18x \\
 & 9.0 & 437.31 & 149.21 & 2.93x & 365.36 & 78.87 & 4.63x & 933.25 & 375.25 & 2.49x & 824.20 & 269.82 & 3.05x \\
 & 10.0 & 499.20 & 194.55 & 2.57x & 428.11 & 122.95 & 3.48x & 1055.12 & 484.69 & 2.18x & 960.63 & 380.62 & 2.52x \\
 & 11.0 & 570.28 & 227.11 & 2.51x & 498.81 & 157.68 & 3.16x & 1194.28 & 574.55 & 2.08x & 1097.18 & 488.82 & 2.24x \\
 & 12.0 & 696.55 & 251.62 & 2.77x & 624.73 & 181.55 & 3.44x & 1466.06 & 634.68 & 2.31x & 1380.21 & 543.45 & 2.54x \\
 & 13.0 & 743.44 & 295.53 & 2.52x & 671.54 & 225.13 & 2.98x & 1576.58 & 728.12 & 2.17x & 1486.42 & 636.20 & 2.34x \\
 & 14.0 & 849.89 & 340.96 & 2.49x & 778.40 & 271.00 & 2.87x & 1808.74 & 849.81 & 2.13x & 1725.38 & 769.93 & 2.24x \\
 & 15.0 & 978.17 & 385.28 & 2.54x & 906.51 & 315.31 & 2.87x & 2056.54 & 951.79 & 2.16x & 1972.35 & 871.37 & 2.26x \\
 & 16.0 & 1040.73 & 416.90 & 2.50x & 968.50 & 346.72 & 2.79x & 2208.27 & 1050.98 & 2.10x & 2137.44 & 971.49 & 2.20x \\
\bottomrule
\end{tabular}
}

\label{tab:perf-under-failure}
\end{table*}

\end{document}